# Evidence for Fully Gapped Strong Coupling S-wave Superconductivity in Bi$_4$O$_4$S$_3$

Shruti, P. Srivastava and S. Patnaik

School of Physical Sciences, Jawaharlal Nehru University, New Delhi-110067, India.

e-mail: spatnaik@mail.jnu.ac.in

## Abstract

We report on the superconducting gap and pairing symmetry in layered superconductor Bi$_4$O$_4$S$_3$. The measurement of temperature dependence of magnetic penetration depth was carried out using tunnel diode oscillator technique. It is observed that Bi$_4$O$_4$S$_3$ is a conventional s-wave type superconductor with fully developed gap. The zero-temperature value of the superconducting energy gap $\Delta_0$ was found to be 1.54 meV, corresponding to the ratio $2\Delta_0/k_BT_c$=7.2 which is much higher than the BCS value of 3.53. In the superconducting range, superfluid density is very well described by single gap s- wave model.

**Introduction**

Superconductivity involves macroscopic quantum condensation of paired electrons. Such pairing could originate from a variety of quantum interactions. Thus, when a new superconducting system is discovered, the first challenge is attained towards deciphering the interaction leading to the pairing mechanism. In this context, the recent discovery of superconductivity in layered sulphide $Bi_4O_4S_3$ by Mizuguchi et al. [1] has once again reignited the field of superconductivity. Its transition temperature ($T_{c0}$) is ~4.5K [1,2]. With high pressure synthesis, a substantial increase in $T_c$ ~10.6 K has been observed for $LaOBiS_2$ [3]. A series of $ReOBiS_2$ compounds have been reported with constituent $BiS_2$ conducting layers [4,5]. Theoretical works based on the first principle band structure calculations [1,6] predict that the dominating bands for the electron conduction as well as for the superconductivity are derived from the Bi $6p_x$ and $6p_y$ orbitals where the electrons typically exhibit weak correlation effects. Interestingly, the structure of these sulfides comprising of $BiS_2$ layers is similar to the layered cuprates and pnictide superconductors. In $Bi_4O_4S_3$, it is the $BiS_2$ layer where superconductivity is achieved by carrier doping either by substitution [3] or by creating deficiency [2].

There have been a variety of reports on the origin of superconducting mechanism in $BiS_2$ based compounds. While nesting of the Fermi surface and large electron phonon coupling constant are predicted for $LaO_{1-x}F_xBiS_2$ [7,8], there are some indication that even for $Bi_4O_4S_3$, the coupling strength could exceed the conventional BCS prediction [9]. There are theories that predict structural instabilities [7,8] in close proximity to competing ferroelectric and charge density wave (CDW) phases [8]. This instability is possibly due to anharmonic potential of the S ions lying on the same plane as the Bi ions. In parallel to the spin-density wave (SDW) instability in Fe-based superconductors, a CDW instability from negative phonon modes at or around the M point, ($\pi$, $\pi$), is suggested to be essential to the superconductivity in LaO$_{1-x}$F$_x$BiS$_2$ [8]. Thus, theoretically, spin fluctuation may also account for the superconducting pairing in these compounds.

Towards this end, the magnetic penetration depth is very important for the determination of the magnitude of the superconducting order parameter and pairing symmetry. In conventional BCS theory, this pairing is attained between two electrons with opposite momenta through exchange of phonons. In the weak electron – phonon coupling scenario, a constant ratio of superconducting gap and superconducting transition temperature ($2\Delta_0 / k_B T_c = 3.53$) is predicted. This can be accessed through the penetration depth $\lambda(T)$ measurements as it is directly related to the superfluid density $\rho \propto 1/\lambda^2$ and hence to the magnitude of the superconducting order parameter $\Delta(T)$ which is the activation energy for quasi-particle excitations of the superconducting ground state and equivalently termed as the superconducting gap. It is an extremely useful probe for distinguishing between different possible superconducting gap structures on the Fermi surface and to reflect on the superconducting pairing mechanism. Several examples are pertinent. Temperature variation in the superfluid density consistent with weak-coupled d wave superconductor has been observed in hole doped cuprates such as YBa$_2$CuO$_{7-\delta}$ [10], Bi$_2$Sr$_2$CaCu$_2$O$_{8+\delta}$ [11]. Quadratic power law dependence of superfluid density on temperature consistent with a d-wave pairing state exhibiting unitary limit impurity scattering was confirmed in Pr$_{2-x}$Ce$_x$CuO$_{4-y}$ (PCCO) [12]. $d$-wave pairing with $\Delta\lambda \propto T^{1.5}$ was observed in organic superconductors κ-(ET)$_2$Cu[N(CN)$_2$]Br [13] and such dependence was also seen in heavy fermion compound CeCoIn$_5$ [14]. Similarly various experiments have provided quantitative evidence for the two-gap nature of superconductivity in MgB$_2$ [15]. It was found that the smaller gap is on the $\pi$ sheet with $\Delta^s$=2.5meV ≈ 0.42$\Delta_{BCS}$ and larger gap is on $\sigma$ sheets with $\Delta^l$=6.45meV ≈ 1.1$\Delta_{BCS}$ ($\Delta_{BCS}$ represents weak coupling BCS value) [15]. In CaAlSi, an anisotrpic 3D s-wave superconductivity was observed [16]. Further, superconducting state in oxypnictides has been found to be consistent with the fully gapped s± pairing symmetry[17]. Some other techniques such as Angle-Resolved Photoemission Spectroscopy (ARPES) [18], Scanning Tunnelling Spectroscopy(STS)[19], Muon Spin Rotation and Mutual Coil Induction [20] have also been used to ascertain pairing symmetry, but these techniques are extremely sensitive to the surface state. The tunnel diode oscillator technique that we have used, is also surface sensitive but it is less susceptible to impurities since the length scale probed is much larger (for type – II superconductors) than the coherence length as probed in a STS. Also because of limited directional resolution it is difficult to use the other techniques in three dimensional anisotropic Fermi surfaces. The magnetic penetration depth on the other hand is a bulk probe where a high degree of resolution can be achieved.

In this paper, we report a detailed penetration depth study on polycrystalline samples of recently discovered superconductor $Bi_4O_4S_3$. We have successfully synthesized this new superconductor with enhanced superconducting transition temperature ($T_{conset}$ ~ 5.3K). The low temperature data of penetration depth measurement were fitted to the BCS equation that yielded a gap ratio $2\Delta_0 / k_B T_c$ = 6.7 which is larger than BCS value of 3.53. Moreover, the gap obtained from the fitting of superfluid density in full temperature range (upto $T_c$) is 7.2 which is almost twice of the weak coupling BCS value suggesting surprisingly strong coupling in $Bi_4O_4S_3$ superconductor.

**Experimental**

Polycrystalline sample of $Bi_4O_4S_3$ was prepared using a conventional solid state reaction method. $Bi_2S_3$, $Bi_2O_3$ powders and S grains were ground, pelletized, sealed into an evacuated quartz tube and heated at 510 ºC for 12 h. The product was well-ground, pelletized and again annealed at 510 ºC for 10 h. Transport and magnetic measurements were undertaken in a *Cryogenic* low temperature and high magnetic field Cryogen Free system in conjunction with variable temperature insert.

The London penetration measurement was carried out using tunnel diode resonator technique which consisted of a copper coil connected parallel to a capacitor and driven to resonance due to a tunnel diode oscillator at radio frequency (2.3MHz at room temperature). The sample was placed inside in this copper solenoid coil which formed the part of the inductor. The amplitude of the rf field generated by coil is ~1 µT which is much less than the lower critical field $H_{c1}$ of $BiS_2$ based superconductors [2]. Hence the sample was in Meissner state in zero applied external dc field. The most important feature of this measurement is the stability of the oscillator. With a proper choice of circuit elements and by taking care of shielding, we were able to obtain a stability of one part in $10^7$ Hz. An oven stabilized frequency counter was used for measurements of frequency. The sample temperature was measured and regulated with a 340 Lakeshore temperature controller. The sample was kept inside the inductor such that the plane of the platelet was perpendicular to the axis of the inductor that formed a part of LC component of the ultra-stable oscillator. A change in the magnetic state of the sample results in a change in the inductance of the coil which is reflected as a shift in oscillator frequency that was measured using an Agilent 53131A counter. Moreover, the shift in the resonant frequency $(F_s - F_0)/F_0$ is proportional to sample linear susceptibility $\chi$ ($F$ is the frequency in the presence of superconducting sample and $F_0$ is the resonance frequency in the absence of a sample) and for a sample in perpendicular to rf magnetic field, the susceptibility is given by,

$$-4\pi\chi = \frac{1}{1-N}[1 - \frac{\lambda}{R}\tanh(\frac{R}{\lambda})] \tag{1}$$

where $R$ the effective sample dimension and $N$ is is the effective demagnetization factor [21]

For penetration depth λ << R as in the present case;

$$\frac{\Delta F}{F_0} = \frac{V_S}{2V_0(1-N)}(1-\frac{\lambda}{R}) \qquad (2)$$

where $V_s$ and $V_0$ are the sample and effective coil volume respectively. Thus a change in the penetration depth $\Delta\lambda(T) = \lambda(T) - \lambda(T_{min})$ from the lowest temperature $T_{min}$ is given by the shift in oscillator frequency as $\Delta\lambda(T) = G\delta F(T)$. Here, $G = RV_s F_0 /[2V_0(1-N)]$ is the calibration constant of the set up that depends on sample and coil geometries and demagnetization factor and $\delta F \equiv \Delta F(T) - \Delta F(T_{min})$ [21]. We found that the value of $G$ estimated from this technique for pure niobium platelet (of almost identical shape and size of our $Bi_4O_4S_3$ specimen) overestimates the change in $\Delta\lambda(T)$. The possible source of error may be in determining the effective coil volume and demagnetization factor. Instead we followed the procedure where the zero field frequency shift was fitted with the temperature dependence of BCS s-wave equation. Taking $\Delta\lambda(T) = G\delta F(T)$, the best fit for the experimental data yielded $G$=2.27 Å/Hz for our set up.

**Results and Discussion**

The Rietveld refined room-temperature XRD patterns for $Bi_4O_4S_3$ sample is reported elsewhere [22] that confirms the I4/mmm phase purity of the specimen. Figure 1 shows the resistivity behaviour near transition temperature for $Bi_4O_4S_3$ and the inset shows dependence of resistivity from 300 K to 2 K for the same. The resistivity varies in a $T^2$ dependence as appropriate for a Fermi liquid state within a narrow window of 25 K to 50 K. The resistivity behaviour for the undoped sample $Bi_6O_8S_5$ on the other hand, exhibits semiconducting behaviour down to the lowest measured temperature [22]. Superconductivity is optimized for 50% deficiency in SO$_4$ concentration that is the equivalent Bi$_4$O$_4$S$_3$ phase.

The inset of Figure 2 shows the change of the penetration depth $\Delta\lambda(T)$ down to 1.68K, and reveals a sharp superconducting transition at T$_c$ ~ 5.0 K . This is slightly lower than onset of resistive transition. The main panel of Figure 2 plots the variation in penetration depth as a function of reduced temperature. In the following, we discuss the low temperature behavior of $\Delta\lambda = \lambda(T) - \lambda(0)$ and the superfluid density, $\rho_s(T) = [\lambda(0)/\lambda(T)]^2$, towards determining the symmetry and magnitude of gap for the superconducting state[23]. In the BCS theory, for a fully gapped s-wave superconductor, the behavior of $\lambda(T)$ asymptotically approaches to

$$\Delta\lambda(T) = \lambda(0)\sqrt{\frac{\pi\Delta_0}{2k_BT}} \exp(\frac{-\Delta_0}{k_BT}), \qquad (3)$$

at low temperature. Here λ(0) and Δ$_0$ are the values of penetration depth $\lambda$ and the superconducting energy gap Δ at *T*=0. This exponential dependence is fairly applicable to the for *T*/*T*$_c$ ≤ 0.5 [24]. In case of a nodeless anisotropic energy gap or distinct gaps on different Fermi-surface sheets, $\lambda(T)$ follows the exponential law but in that case, Δ$_0$ approximately equals to the minimum energy gap in the system, and $\lambda(0)$ gives effective value that depends on the details of the gap anisotropy [23]. For *d*-wave pairing in the clean limit, on the other hand,

$$\Delta\lambda(T) \approx \lambda(0)\frac{2\ln 2}{\alpha\Delta_0}T \qquad (4)$$

where $\alpha = \Delta_0^{-1}[d\Delta(\Phi)/d\Phi]_{\Phi\to\Phi_{node}}$ and Δ(Φ) is the angle dependent gap function [23]. In the presence of impurity scattering, *d*-wave gap is suppressed and the temperature dependence changes from linear behaviour to power law; $\Delta\lambda \sim T^2$ [25].

Qualitatively, the observed weak temperature dependence below 2.5 K in Figure 2 is consistent with s-wave BCS theory. This also marks the signature of a fully developed superconducting gap. Fitting the low temperature data with power law $T^n$ yields a very high value of $n \sim 7$ which rules out the possibility of $d$ wave pairing. At low temperature $T/T_c < 0.5$, the data are best fitted to the BCS Eq.(3) that yields gap ratio $2\Delta_0/k_B T_c = 6.7$ ±0.3 and the corresponding value of energy gap $\Delta_0 \sim 1.43$ meV. This gap ratio is very high compared to the weak coupling limit of 3.53. This suggests that $Bi_4O_4S_3$ is a strongly coupled superconductor. Such possible strong coupling in this material has been made in a detailed theoretical analysis by Yildirim [26]. Evidence for strong coupling is also seen in STS measurement [9], although the value of gap ratio reported is much higher; $2\Delta_0/k_B T_c \approx 17$. To bring in a prospective, the observed gap value from our experiments can be compared with some of the other strong coupling superconductors such as 4.8 in $KOs_2O_6$ [27,24], 5.2 in $PrOs_4Sb_{12}$ [28], 4.4 in $SrPd_2Ge_2$ [29], 3.87 in $RbOs_2O_6$ [27], ~5 in $SrPt_3P$ [30], ~ 8 in $PuCoGa_5$ [31] and 7.74 in $YBa_2Cu_3O_{6.98}$ [32].

A s-wave pairing with sign reversal is suggested if spin fluctuations play main role in cooper pairing [6]. With regard to $Bi_4O_4S_3$, Y.Gio has formulated a theory base on two orbital model for the pairing symmetry on the basis of three possibilities, the isotropic s-wave, the anisotropic s-wave and the $d$-wave pairing [33]. To analyze our data in the light of these possibilities, in Figure 3 we plot the normalized superfluid density, $\rho(T) = \left[\lambda(0)/\lambda(T)\right]^2$ as a function of reduced temperature over full temperature range. Best fits to the superfluid density over the entire temperature range allows one to detect the existence of multiple gaps as well as their temperature dependences [34], or the anisotropies of the superconducting gap(s). But this method requires $\lambda(0)$, the zero temperature London penetration depth data, that are not available for $Bi_4O_4S_3$. We estimated the value of $\lambda(0)$ from lower critical field

$$H_{c1} = \frac{\Phi_0}{4\pi\lambda^2}\left(\frac{\ln\lambda}{\xi}+0.485\right) \tag{5}$$

where the value of $H_{c1}(0)$ and $\xi(0)$ are given by ~ 15 Oe and 110 Å respectively [2, 22], and $\Phi_0$ is magnetic flux quantum = $2 \times 10^{-7}$ G cm$^2$. The value of $\lambda(0)$ as calculated from the above formula is ~734nm. Further, in the dirty limit case $(l<\xi)$ and in the clean limit $(l>\xi)$, where $l$ is mean free path, the super fluid density $\rho(T)$ has the form [36],

$$\left.\frac{\lambda^2(0)}{\lambda^2(T)}\right|_{dirty} = \frac{\Delta(T)}{\Delta_0}\tanh\left[\frac{\Delta(T)}{2k_B T}\right], \tag{6}$$

$$\left.\frac{\lambda^2(0)}{\lambda^2(T)}\right|_{clean} = 1+2\int_{\Delta(T)}^{\infty}\left[\frac{\partial f}{\partial E}\right]\frac{E}{\sqrt{E^2-\Delta(T)^2}}dE, \tag{7}$$

where $f=[1+\exp(E/k_B T)]^{-1}$ is the Fermi function with temperature dependent gap function,

$$\Delta(T) = \Delta_0 \tanh\left[\frac{\pi k_B T}{\Delta_0}\sqrt{a\left(\frac{T_c}{T}-1\right)}\right] \tag{8}$$

Here $\Delta_0$ is the gap magnitude at zero temperature and '$a$' is a free parameter that depends on the particular pairing state and $k_B$ is Boltzmann constant. The integration is over the entire quasi-particle energies measured from the chemical potential [23]. As shown in Figure 3, for the estimated value of $\lambda(0)=$ 734nm a comparison is attempted for the two fluid model [35], and the clean and dirty s-wave models [36]. Evidently, the superfluid density for Bi$_4$O$_4$S$_3$ does not fit at all to the two weak coupling BCS s–wave curves corresponding to dirty and clean limits and neither do the data follow the two fluid Gorter and Casimir model. Subsequently, the $\rho(T)$ data were fitted using eq. (7) and gap function (8) taking '$\Delta_0$' and '$a$' as free parameter using *Mathematica*. As is evident, the superfluid density data show excellent fitting for a single s-wave equation for $2\Delta_0/k_B T_c= 7.18 \pm 0.17$ with the corresponding energy gap value $\Delta_0 \sim$ 1.54 meV and $a = 1.23 \pm 0.007$. This value of gap ratio

$2\Delta_0 = 7.2k_BT_c$ is almost double of weak coupling BCS value of 3.53 which again underlines the strong electron – phonon coupling in $Bi_4O_4S_3$. It should be noted that the gap values estimated from raw fitting of BCS equation would be less accurate than the gap estimated from the full temperature range super fluid density since the BCS equation is applicable at asymptotically approached lowest temperature. The value of 'a' however is higher than its value for isotropic s-wave gap (a ~ 1) for weak coupling. The value of 'a' relates to the specific heat transition [23]. Since we don't have specific heat data, no conclusive comment can be made on the obtained best fit value of 'a'. An attempt to ascertain two gap fitting using *alfa* model for two gap superconductivity [15], was also undertaken. The technique involved letting the values of constants get determined self consistently, and the results indicated convergence to a single gap that convincingly exclude the possibility of two gaps in the present case.

**Conclusion**

In conclusion, a study of temperature dependent magnetic penetration depth and super fluid density in the newly discovered layered sulphide $Bi_4O_4S_3$ superconductor confirms a fully gapped s-wave pairing in the strong coupling limit. The low temperature $\lambda(T)$ values show exponential temperature dependence. $\Delta\lambda$ tends to be rather flat below $0.5T_c$ and marks the signature of fully developed superconducting gap. Superfluid density is very best fitted with single gap s-wave equation with gap ratio $2\Delta_0=7.2k_BT_c$ that indicates the existence of strong electron phonon coupling in this primary $BiS_2$ based layered superconductor.

**Acknowledgment**

Shruti acknowledge UGC for providing her the JRF fellowship. PS is grateful to UGC, India for the award of a Dr D S Kothari postdoctoral fellowship. Technical support from AIRF is gratefully acknowledged. SP acknowledges support under DST – FIST program of Govt. of India.

**Figure Captions:**

**Figure 1.** Temperature dependence of resistivity showing superconducting transition temperature of 5.3K. Inset shows metallic behaviour of resistivity upto room temperature.

**Figure 2.** Change in penetration depth $\Delta\lambda(T)$ showing exponential fitting(red line) using Eq.(3). Inset shows temperature dependence of normalized $\Delta\lambda(T)$ with superconducting transition at 5.0K .

**Figure 3.** Comparison of temperature dependence of superfluid density of $Bi_4O_4S_3$ with clean s-wave, dirty s-wave and two fluid models. Red line shows the best fitting for a single s-wave equation with fitting parameter $2\Delta_0/k_BT_c$= 7.18 ± 0.17 and $a$ = 1.230 ± 0.007.

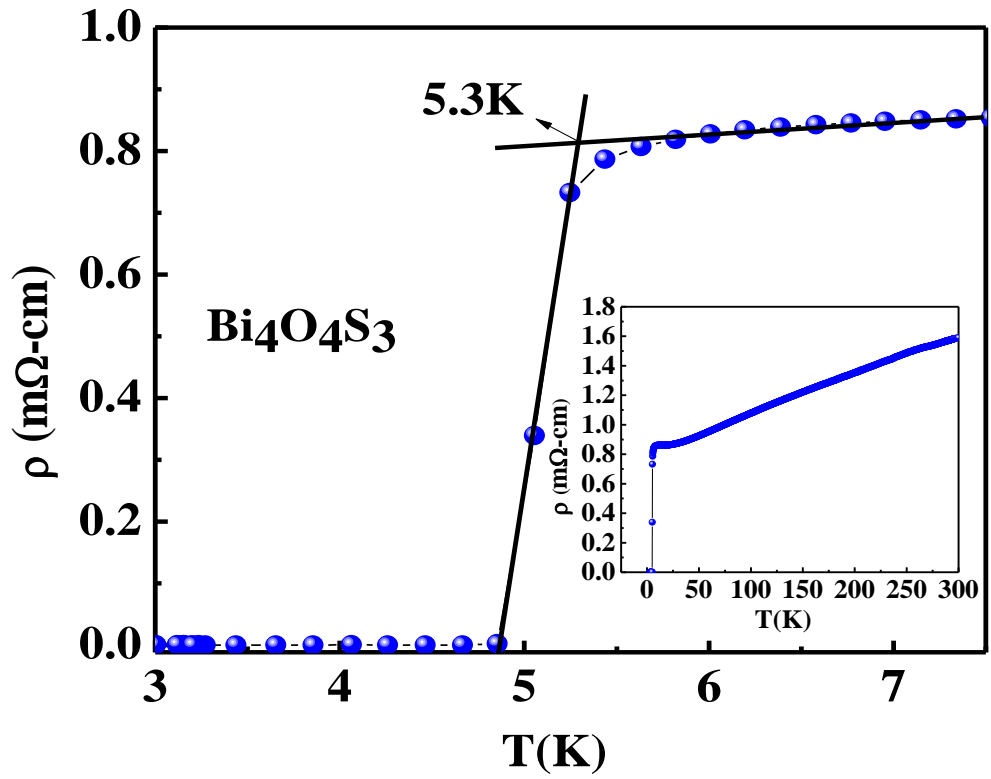

Figure1.

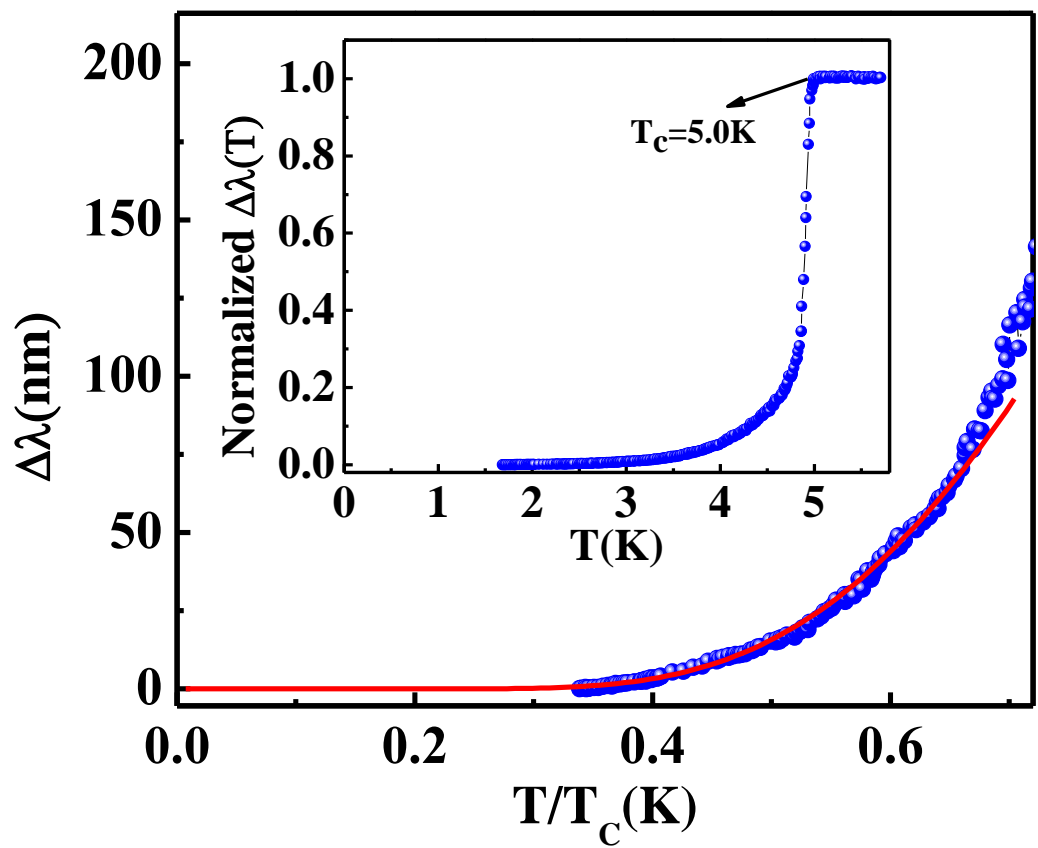

**Figure 2.**

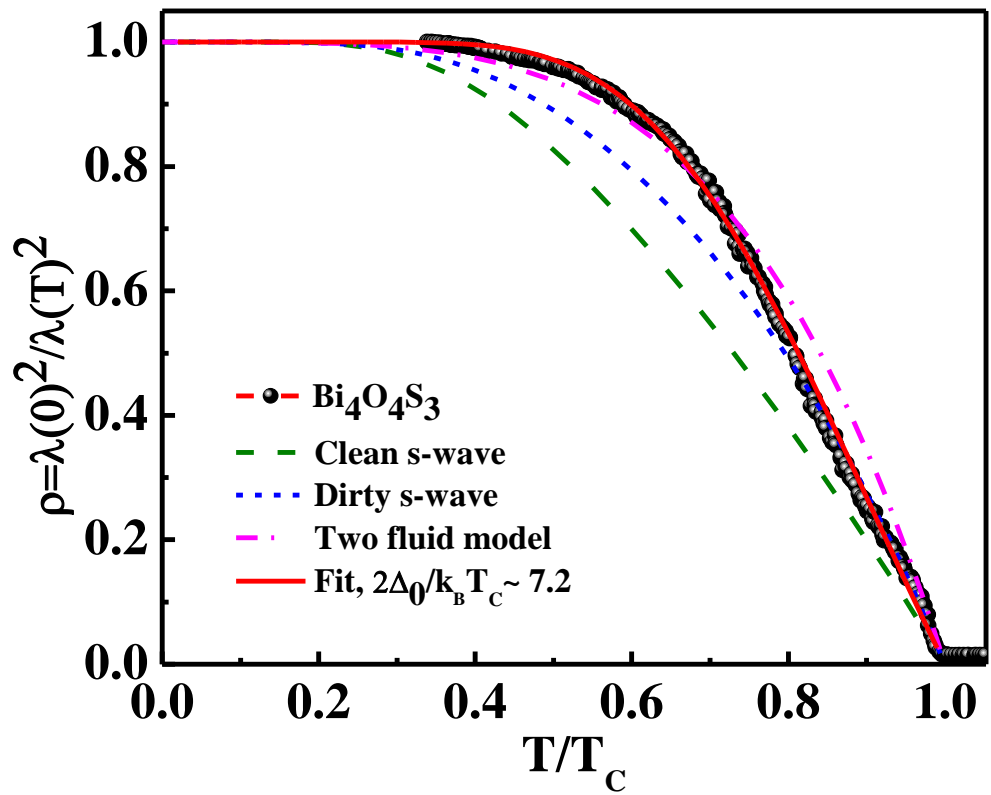

**Figure 3.**